\documentclass[fleqn,10pt]{wlscirep}
\title{Conformal Vortex Crystals}

\usepackage{hyperref}
\usepackage{rotating}
\usepackage{graphicx}
\usepackage{latexsym}
\usepackage{amssymb}
\usepackage{amsfonts}
\usepackage{soul}
\usepackage{color}
\usepackage{amsmath}
\usepackage{bm}
\usepackage{lipsum}
\usepackage{longtable}

\author[1]{Ra\'{\i} M. Menezes}
\author[1,*]{Cl\'ecio C. de Souza Silva}
\affil[1]{Departamento de F{\'{\i}}sica, Universidade Federal de Pernambuco, Cidade Universit\'aria, 50670-901 Recife-PE, Brazil}
\affil[*]{clecio@df.ufpe.br}

\date{\today}

\begin{abstract}
\textbf{
We investigate theoretically globally nonuniform configurations of quantized-flux vortices in clean superconductors trapped by an external force field that induces a nonuniform vortex density profile. Using an extensive series of numerical simulations, we demonstrate that, for suitable choices of the force field, and bellow a certain transition temperature, the vortex system self-organizes into highly inhomogeneous conformal crystals in a way as to minimize the total energy. These nonuniform structures are topologically ordered and can be mathematically mapped into a triangular Abrikosov lattice via a conformal transformation. Above the crystallization temperature, the conformal vortex crystal becomes unstable and gives place to a nonuniform polycrystalline structure. We propose a simple method to engineer the potential energy profile necessary for the observation of conformal crystals of vortices, which can also be applied to other 2D particle systems, and suggest possible experiments in which conformal or quasi-conformal vortex crystals could be observed in bulk superconductors and in thin films. 
}
\end{abstract}

\begin{document}

\flushbottom
\maketitle
%
%
\thispagestyle{empty}

\section{Introduction}

Six decades after its discovery, the uniform Abrikosov lattice~\cite{Abrikosov1957} remains as the only known ordered state of quantized vortices in type-II superconductors. Hexagonal~\cite{Essmann1967,Trauble1968a,Hess1989,Bishop1992,Eskildsen2002}, and in more rare instances even deformed hexagonal and square vortex lattices~\cite{Wilde1997,Riseman1998,Sosolik2003} are typically found in clean superconductors cooled under an applied magnetic field. In contrast, when a type-II material is cooled at zero field and only then an external field is applied, the final vortex distribution (called critical state) is typically nonuniform and highly disordered as a result of the balance between the incoming flux gradient and the random pinning forces produced by material inhomogeneities~\cite{Bean1962}. Recently, the critical state in superconducting films with periodic and graded arrays of pinning centers has been investigated in detail~\cite{Silhanek2011,Misko2012,Motta2013,Ray2013,Wang2013,Guenon2013,Ray2014,Wang2016}. In particular, graded pinning arrays constructed via a conformal transformation present remarkable vortex pinning properties~\cite{Misko2012,Ray2013,Wang2013,Guenon2013,Ray2014,Wang2016}. 
However, notwithstanding the underlying order of these pinning arrays, the nonuniform vortex configurations reported so far follow the pinning distribution only locally, and the global order of the vortex arrangement in these systems remains unclear~\cite{Misko2012,Ray2014}. The more fundamental problem of whether nonuniform vortex crystals can grow in a self-organized manner, that is, without any underlying ordered structure, has not been tackled so far.



The problem of how two-dimensional many-body systems self-organize in order to cope with an imposed density inhomogeneity or surface curvature has inspired growing interest~\cite{Bausch2003,Irvine2010,Koulakov1998,Mughal2007,Yao2013,Cerkaski2015,Terrones2010,Sanjuan2014,Negri2015}. A common feature in these systems is the presence of topological defects, appearing either isolated or in groups, that provide the necessary bending of lattice lines while frustrating the global orientational order~\cite{Koulakov1998,Cabral2004,Mughal2007,Yao2013,Cerkaski2015,Bausch2003,Irvine2010,Azadi2014,Azadi2016}. 
However, for a very limited class of planar 2D systems, namely magnetized spheres compressed by gravity~\cite{Pieranski1989,Rothen1993} and confined foams~\cite{Drenckhan2004}, the particles were observed to self-organize into a highly ordered, quasi-conformal crystal with a nonuniform density profile. Remarkably, these structures, dubbed gravity rainbow, can be approximately mapped into a uniform hexagonal lattice via a conformal transformation. 


Mathematically, a conformal point lattice in the complex plane $z$ (representing the real $x$-$y$ plane) is the result of mapping a regular, say hexagonal, lattice defined in an auxiliary plane $w$ via a conformal transformation $z(w)=x(u,v)+iy(u,v)$. Since, by definition, such a transformation preserves angles, the lattice in $z$ inherits the local hexagonal symmetry of the original one. In addition, the transformation results in a unique, generally nonuniform density distribution given by $n_z = \left|\frac{dw}{dz}\right|^2n_w$, with $n_w=\text{const.}$ representing the uniform distribution of lattice points in $w$. 
Therefore, the idea of subjecting particles to a confining force field capable of inducing a conformal density profile looks a very promising way of discovering new globally nonuniform ordered structures in planar systems~\cite{Pieranski1989,Rothen1993,Drenckhan2004}. However, finding such force field does not guarantee a conformal crystal as, in general, the actual physical problem cannot be reduced to a coordinate transformation. The system may instead follow the same density profile in e.g. a disordered (glassy) or segmented (polycrystalline) way~\cite{Wojciechowski1996}. Indeed, it remains unknown whether conformal crystals provide the minimum energy of interacting particles under a straining force field and whether such structures are stable with respect to the thermal excitation of topological defects.




In this article, we demonstrate that a vortex system subjected to a suitable force field crystalizes spontaneously in a topologically ordered, conformal lattice. The external force field was previously calculated analytically within a simple continuum model by constraining the density profile to follow that of a perfect conformal lattice for two different situations: (i) bulk superconductors, where vortex-vortex interactions are short range, and (ii) thin films, where interactions are long-range. Following a thorough series of numerical annealing processes, we show that, in both cases, the conformal vortex crystal (CVC) phase provides the minimum energy at zero temperature and that it remains stable with respect to the proliferation of topological defects up to a finite temperature $T_{CVC}$. Above this temperature, the CVC gives place to a polycrystalline phase, comprising randomly oriented conformal crystallites, before finally melting at a higher temperature.  Since the analytical model applies for any given interaction potential, including long-range interacting systems, we expect that the results presented here can be relevant to other confined systems of interacting objects, e.g. colloids, plasmas and vortices in Bose-Einstein condensates.

\section*{Results}

\subsection*{Continuum limit analysis} 
Our first step towards a conformal vortex crystal (CVC) is to find what ideal external potential can accommodate particles, in general, or vortices, in particular, in a strictly conformal density profile. In the literature, estimates of the confining potential are restricted to systems of particles interacting via inverse power laws with exponents $k>2$~\cite{Rothen1996}. This class of interaction potential excludes a broad range of systems of interest, such as plasmas, colloids and superconducting vortices. Here, we approach this problem within the continuum approximation, which assumes that inter-particle spacings are smaller than any other length scale in the system. In this limit, the free-energy of a system of $N$ particles interacting via a pair potential $V_\text{int}(\bm{r},\bm{r}')$ and subjected to an external potential $U(\bm{r})$ can be expressed as a functional of the particle distribution:
%
$
   {\cal F}[n(\bm{r})] = \int d\bm{r} \,n(\bm{r})\,U(\bm{r}) + 
   \frac{1}{2}\int d\bm{r}\,d\bm{r}' \,n(\bm{r})V_\text{int}(\bm{r},\bm{r}')n(\bm{r}').
$ 
%
By minimizing ${\cal F}$ with respect to $n(\bm{r})$, one finds the non-local balance relation
\begin{equation}\label{eq.U}
   U(\bm{r}) = -\int d\bm{r}' \,n(\bm{r}')\,V_\text{int}(\bm{r},\bm{r}') + C,
\end{equation}
where $C$ is a constant to be determined from the number conservation condition $\int d\bm{r}\,n(\bm{r})=N$. 
Notice that equation~\eqref{eq.U} can be approximated by $U(\bm{r}) = -g(\bm{r}) n(\bm{r})$, with $g(\bm{r})=\int d\bm{r}' \,V_\text{int}(\bm{r},\bm{r}')$, for the cases where $n(\bm{r})$ changes on a scale much larger than the characteristic length of the interaction potential. If, in addition, the interparticle force law is central, $g$ becomes a constant and $n(\bm{r})$ is essentially the negative copy of the potential. This very simple result, which we shall refer to as local approximation, applies only when the interaction potential is short-ranged, as is the case of the vortex-vortex potential in a bulk superconductor. Otherwise, the full non-local character of Eq.~\eqref{eq.U} must be dealt with. 

To be specific, we shall henceforth focus on one-dimensional external potentials, $U(y)$. In this case, the logarithmic map, $z(w)=-i\ell\ln (iw/\ell)$, is the only transformation that produces a one-dimensional density profile~\cite{Rothen1996,Wojciechowski1996}. Therefore, the required conformal density distribution in the physical $z$ plane is $n(y)=n_0e^{-2y/\ell}$, where $n_0$ is given by the number conservation condition. In addition, to deal with a finite number of vortices, we assume that the potential is periodic on a length $L$, that is $U(y+L)=U(y)$, and periodic boundary conditions in the $x$ direction, also over a length $L$. We shall also fix $\ell=L/\pi$, so that the prescribed conformal transformation maps a rectangle of base $L$ into a semiannular region. In this case, since $\ell\ll2L$, $n_0=2N/[\ell(1-e^{-2L/\ell})]\simeq2N/\ell$. Results for other values of $\ell$ are presented and discussed in the Supplementary Information available online.

It is worth mentioning that Eq.~\eqref{eq.U} is valid for any kind of monotonous pair potential, covering both short and long range interactions, as long as interparticle spacings are smaller than the characteristic length scales of both the interaction potential and the confining potential. As such, it generates the potential energy capable of inducing a coarse-grained particle distribution with any desired profile, including the conformal profile. However, it does not guarantee what the detailed structure of the particle distribution will be. For that, thorough minimization of the full many-body problem is needed. 

\subsection*{Vortices in bulk superconductors} 

In a bulk superconductor of thickness $d\gg\lambda$, where $\lambda$ is the London penetration depth, vortices interact with each other via a central pair potencial given by $V_{\rm int}(r)=\epsilon_0 K_0(r/\lambda)$, where $K_0(z)$ is the zeroth-order modified Bessel function of the second kind and $\epsilon_0=\phi_0^2d/2\pi\mu_0\lambda^2$ (with $\phi_0$ the flux quantum and $\mu_0$ the vacuum permeability). For large distances, this interaction decays exponentially with a characteristic length $\lambda$. Therefore, by further assuming that $\ell\gg\lambda$, the local approximation can be used, and the necessary external potential can be written as  
\begin{equation}\label{eq.U(y)}
   U(y) = 
   \begin{cases}\displaystyle
      -gn_0~y/\tilde\xi, &\quad 0\leq y<\tilde{\xi}, \\
      -gn_0~e^{-2(y-\tilde{\xi})/\ell}, &\quad \tilde{\xi}\leq y\leq L,
   \end{cases}
\end{equation}
where the first term represents a soft-wall that confines the distribution in the $y>0$ region, $g=\phi_0^2/\mu_0$. The wall width $\tilde{\xi}$ was chosen in a way that the maximum force exerted by the external potential, $gn_0/\tilde{\xi}$, is less than the vortex-antivortex unbinding force, thus avoiding violation of vortex conservation.

\begin{figure}[t]
\centering
\includegraphics[width=0.5\columnwidth]{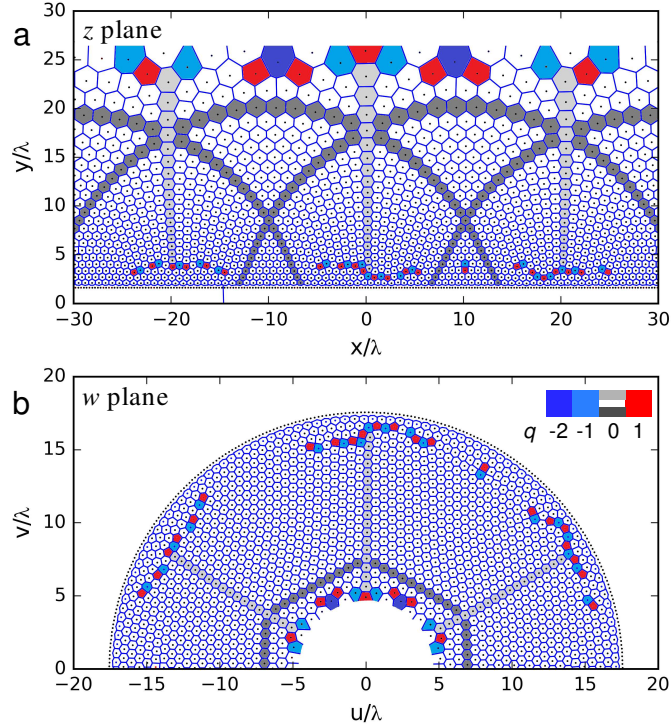}
\caption{Conformal crystal configuration and topological defects. (a) Typical as-annealed low-energy configuration of vortices (represented by dots) in an exponential confining potential and corresponding Voronoi construction (lines). For clarity, we disregarded the lower row of vortices in the Voronoi construction. (b) Inverse conformal map of the configuration in (a) into the auxiliary $w$ plane, which demonstrates the conformal nature of the CVC obtained from the simulations (see text). The face color coding of the polygons depict the local topological charge $q=\nu_0-\nu$, where $\nu$ is the coordination number of the vortex and $\nu_0$ the corresponding value expected for a topologically flat vortex configuration, i.e., $\nu_0=6$ (4), for a vortex in the system bulk (edge). The gray shades are guides to the eye for better identification of the arch-pillar structure of the conformal crystal. }
\label{fig:CVC_zw}
\end{figure}

To find low energy configurations of the vortex system, we performed a series of Langevin dynamics simulations of $N=3000$ vortices in a $L\times 2L$ simulation box with periodic boundary conditions and $L=60\lambda$ following a standard simulated annealing scheme, which was repeated over 50 distinct realizations for a representative statistics (see Methods). Fig.~\ref{fig:CVC_zw}-(a) depicts an example of the most common low-energy configuration found. Only a $60\lambda\times30\lambda$ region containing half the number of vortices is shown. The Voronoi construction reveals that the structure is highly ordered, with most vortices having six neighbors. Topological charges, here defined as the discrepancy in the number of first neighbors of a vortex with respect to the perfect sixfold coordination, are present near the upper and lower boundaries. As we shall discuss later these charges play an important role in stabilizing the curvature of the lattice lines. To check how close this configuration is to a perfect conformal lattice, we applied the inverse transformation, $w(z)=-i\ell e^{iz/\ell}$, shown in Fig.~\ref{fig:CVC_zw}-(b). Indeed, the vortex lattice mapped into the $w$ plane is almost perfectly hexagonal and the vertical pillars and arches seen in the $z$ plane appear in the $w$ plane as, respectively, radial lines forming angles of $60^\circ$ and the sides of a regular hexagon. This qualifies the observed structure as a conformal vortex crystal.

\begin{figure}[t]
\centering
\includegraphics[width=0.5\columnwidth]{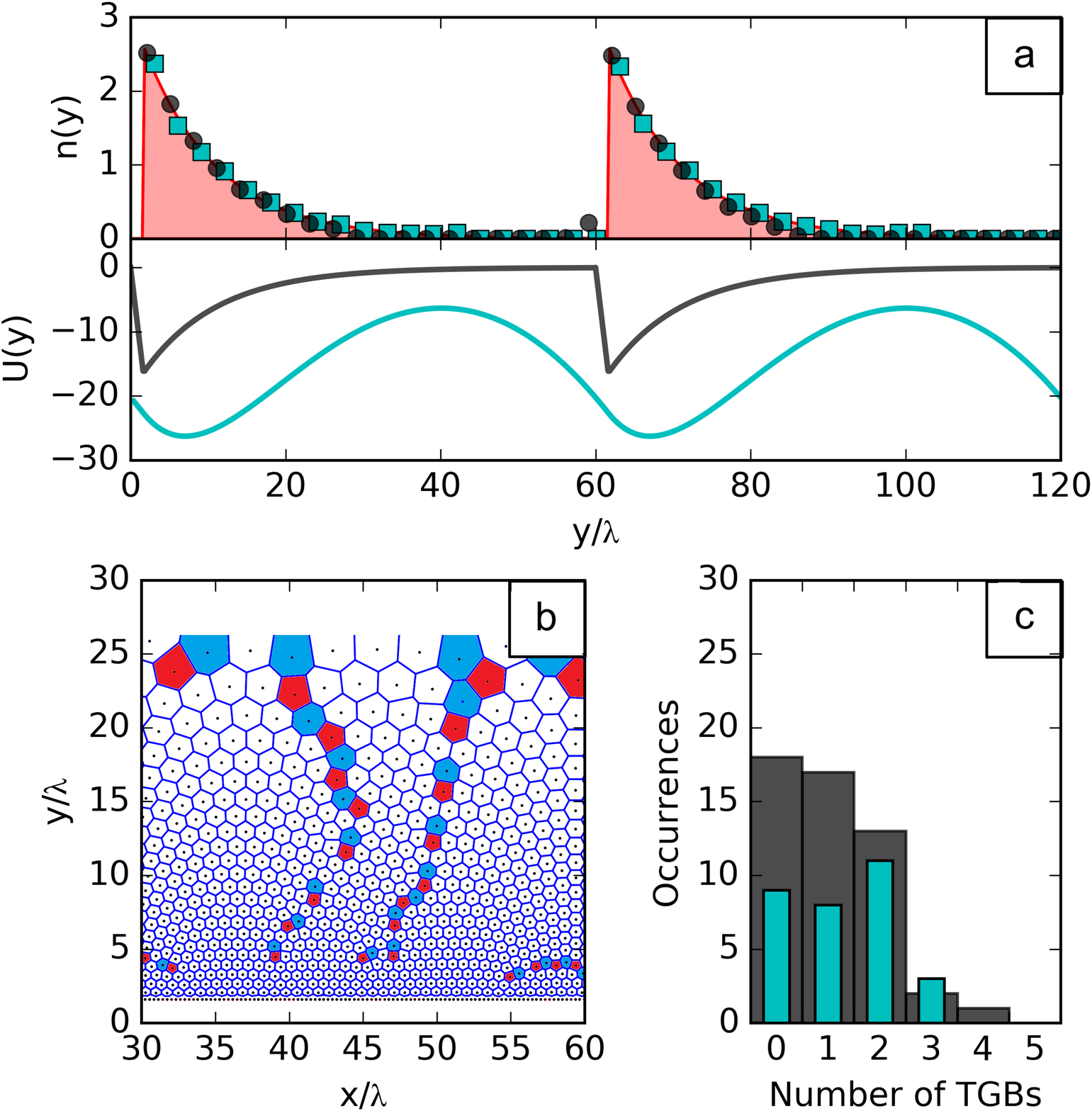}
\caption{Vortex density profiles and classification of the zero-temperature configurations. (a) Top: Vortex density profiles for conformal vortex crystals in bulk samples (circles) and in thin films (squares). The area graph represents the expected exponential profile. Bottom: Potential energy profiles used in the simulations for the bulk (dark gray line) and thin-film (cyan line) cases, see Eq.~\eqref{eq.U}. (b) A typical conformal configuration exhibiting transverse grain boundaries (TGBs). (c) Distribution of the configurations obtained in the simulations for a bulk sample (gray bars) and a thin film (cyan bars) according to the number of TGBs.  }
\label{fig:profile+defects}
\end{figure}

All other observed configurations presented a similar structure, with arches and pillars. However, many of them were found to be broken into domains of conformal crystals separated by prominent, transverse grain boundaries (TGBs), such as those shown in Fig.~\ref{fig:profile+defects}-(b). 
These dislocation lines are topologically neutral and therefore cost little energy, making it difficult to discern which configuration is closest to the ground state of the system. However, by counting the number of TGBs of 50 different realizations of the annealing procedure we can conclude that the single-domain, conformal vortex structure is the most frequent configuration and that the frequency of configurations with more than two TGBs drops fast and represents less than one third of the occurrences [see Fig.~\ref{fig:profile+defects}-(c)].

An important feature, common to all configurations irrespective of the number of TGBs, is that vortices close to the minimum of the external potential tend to form a conventional Abrikosov lattice with a principal axis alined with the $x$ axis. Going up in the $y$ axis by a few vortex rows a transition to the conformal configuration can be identified. In the $w$ plane, this depletion zone is seen as concentric vortex rings near the outer rim. While smooth at some regions, the transition is abrupt just below a pillar, where a sudden $30^\circ$-rotation of the principal axis can be observed. This sharp transition is delimited by high-angle grain boundaries, known as scars, which are typical defect structures found in large 2D particle systems on curved surfaces and are responsible for distributing the necessary curvature in those systems~\cite{Bausch2003,Irvine2010,Negri2015}. Here, the net topological charge of each scar is precisely $+1$ and is counterbalanced by a defect of charge $-1$ near the top of each pillar. Such configuration is responsible for the deformation necessary to accommodate the conformal crystal while keeping the system globally neutral as imposed by the periodic boundary condition along the $x$ axis. A few extra dislocations found at the top of the configuration in $z$ (or inner rim in $w$) are associated to another effect, seen in the $w$ plane as a gentle but progressive dilation of the hexagonal cells as one approaches the inner rim.  This unexpected behavior reflects the failure of the density profile to follow precisely the exponential shape  [see \ref{fig:profile+defects}-(a)], specially in the region $y\geq 15\lambda$, where vortex spacings become larger than $\lambda$ and the continuum approximation breaks down.

\subsection*{Vortices in thin films}

In contrast to the above described situation, vortices in thin films interact via a long range potential, which, for $\Lambda\equiv\lambda^2/d\rightarrow\infty$, is essentially logarithmic, $V(r)=-\epsilon_0\ln r$. In addition, in order to fulfill the periodic boundary conditions, one must take into account the contribution of an infinite set of replicas for each vortex~\cite{Gronbech-Jensen1996}, which results in an effectively non-central interaction. These properties invalidate the local approximation used in the bulk case. Indeed, the calculated configuration of logarithmically interacting vortices in a potential described by Eq.~\ref{eq.U(y)} is non-conformal and characterized by a flat profile (see the Supplementary Information). However, by numerically integrating the non-local relation, Eq.~\eqref{eq.U}, using the logarithmic interaction with the appropriate boundary conditions and performing the simulated annealing scheme over 30 distinct realizations of the random force, conformal vortex crystals with the desired exponential density profile could be observed, see top panel of Fig.~\ref{fig:profile+defects}-(a) (a typical conformal configuration is shown in the Supplementary Information). Although the external potential for accommodating CVCs in thin films is very different from that designed for vortices in bulk samples [see bottom panel of Fig.~\ref{fig:profile+defects}-(a)], their main features are similar. However, some differences are noteworthy. For instance, the distribution of transverse grain boundaries is slightly different [Fig.~\ref{fig:profile+defects}-(c)] and suggests that a perfect conformal crystal is more difficult to achieve in thin films. On the other hand, due to the long range nature of the interactions, the density profile seems to fit better to the exponential shape predicted by the continuum theory. Indeed the mapped configurations depict a more homogeneous distribution near the inner rim in contrast to the bulk sample case.

\subsection*{Thermal excitation of topological defects}

\begin{figure}[tb!]
\centering
\includegraphics[width=0.6\columnwidth]{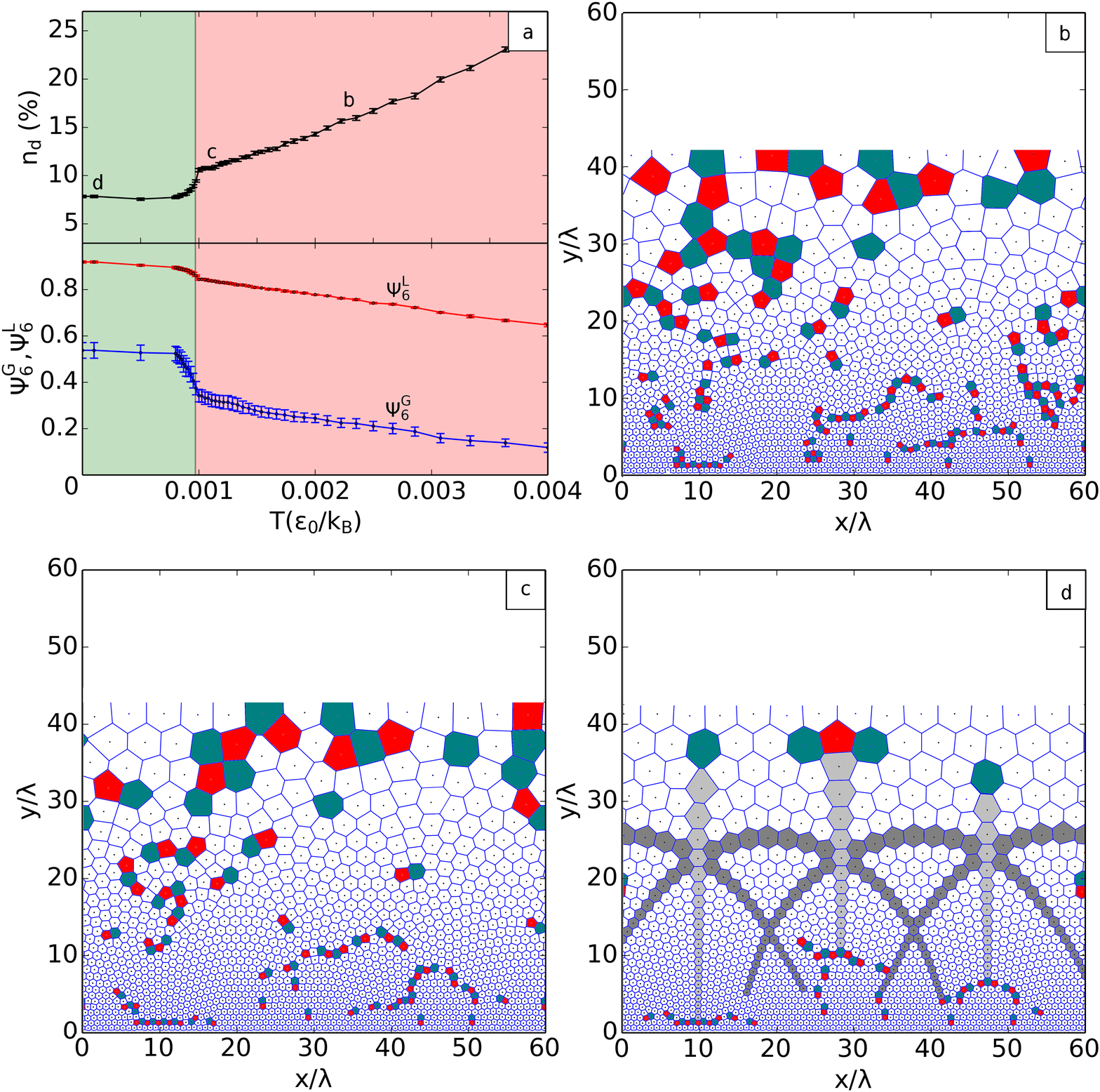}
\caption{Breakdown of the conformal crystal phase into a conformal polycrystal. (a) Temperature dependence of (top) the defect concentration, $n_d$, and (bottom) the local, $\psi_6^L\equiv\langle|\phi_6|\rangle$, and global, $\psi_6^G\equiv|\langle\phi_6\rangle|$, bond-orientational order parameters. Here $\phi_6=\frac{1}{N_b}\sum_{n=1}^{N_b} e^{i6\theta_{mn}}$ measures the mean orientation of the $N_b$ bonds connecting vortex $m$ with its neighbors, $\theta_{mn}$ being the bond angles with respect to a reference axis, and $\langle\cdots\rangle$ means average over all vortices except those at the upper and lower boundaries of the configuration . Error bars represent one standard deviation of the mean over different realizations of the annealing procedure. The abrupt change in the behavior of $n_d(T)$ and $\psi_6^G(T)$ indicate a transition between a globally ordered conformal crystal (green background) and a conformal polycrystalline phase (red background). (b)-(d) Vortex configurations and respective Voronoi constructions at the points specified in (a).}
\label{fig.S4}
\end{figure}

For each temperature, during the annealing processes, we calculated a few important quantities that characterizes the global topological order of the vortex lattice: (i) the density of topological defects, $n_d(T)$, and (ii) the bond-angle orientation order parameters. The defect density allows for a quick determination of the crystallization temperature $T_{\rm CVC}$ of the conformal crystal phase (see Fig.~\ref{fig.S4}). In both cases, bulk ($T_{\rm CVC}=1.6\times10^{-3}\epsilon_0/k_B$) and thin films ($T_{\rm CVC}=0.9\times10^{-3}\epsilon_0/k_B$), the number of defects were observed to stabilize at a low defect concentration from $T_{\rm CVC}$ down to $T=0$ and increase steadily for $T>T_{\rm CVC}$. In Fig.~\ref{fig.S4} we show the detailed results for vortices in thin films. Above the crystallization temperature, thermal fluctuations excite defect pairs and fault lines, such as those seen in other graded vortex distributions~\cite{Trauble1968a,Braun1996}, which break the single CVC structure into small domains (Fig.~\ref{fig.S4} c and d).   

In order to characterize both the local and the global orientational orders, we calculated, respectively, the local, $\psi^L_6(T)$, and the global, $\psi^G_6(T)$, bond-orientational order parameters. Since these quantities are defined for planar (that is, homogeneous) particle systems, we evaluate them in the auxiliary $w$ plane, where the conformal structure is expected to assume the planar hexagonal order. As depicted in Fig.~\ref{fig.S4}, $\psi^L_6$ decreases smoothly as $T$ increases and a small kink is seen around $T_{\rm CVC}$, while $\psi^G_6$ disclose a much more dramatic drop at $T_{\rm CVC}$. These results reveal that the breakdown of the CVC phase is characterized by the loss of global orientational order, induced by extended defect lines forming a polycrystalline structure, while the conformal order within the grains persists up to much higher temperatures, until the system finally melts.

\section*{Discussion}

Let us now briefly discuss the possibility of the experimental realization of a CVC. Thin films offer a more direct approach. Since these materials are unable to screen magnetic fields, one can print flux-density landscapes directly using an external magnetic texture, produced either by permanent magnets or by current-carrying wires conveniently placed on top of the superconducting film. These tools have been explored exhaustively on the mesoscopic scale as a means of manipulating vortices individually~\cite{Morgan1998,Lange2003,deSouzaSilva2007}. In order to observe a conformal or quasi-conformal vortex crystal, one would need to design magnetic textures that resemble a smooth exponential flux decay in a length scale covering many vortex lattice spacings. As an example, we estimate $T_{\rm CVC} =0.37 T_c = 2.5$ K for the thin, plain, MoGe film used in Ref.~\cite{Motta2013}, which renders the CVC phase experimentally accessible in conventional cryogenic setups. 

\begin{figure}[t]
\centering
\includegraphics[width=0.5\columnwidth]{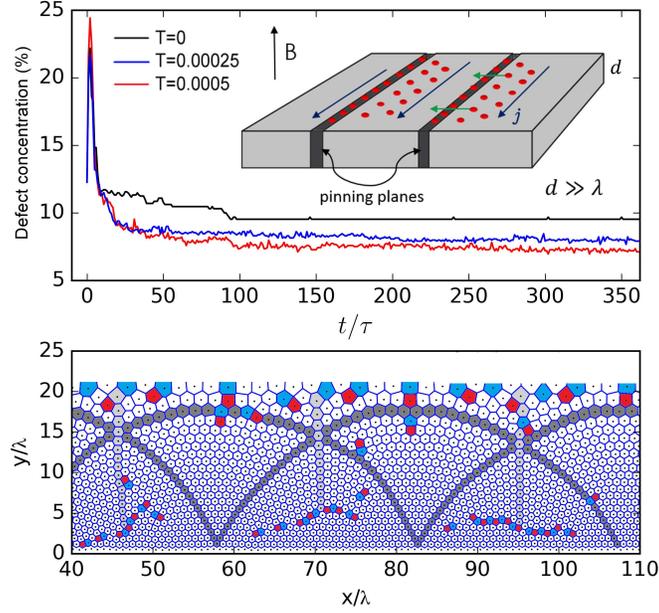}
\caption{Proposal for the experimental observation of a quasi-conformal vortex crystal. Inset: schematic representation of a superconducting crystal containing parallel pinning walls (dark gray regions) where vortices (red dots) are strongly pinning and act as a dam preventing other vortices to pass through. The current density $\bm{j}$ is applied in a way that the induced Lorentz force pushes the vortices in the clean regions against the vortex barriers. Top: concentration of topological defects in the vortex system as a function of time (in units of $\tau=30\eta\lambda^2/\epsilon_0$) after a current density, applied parallel to the pinning walls, is turned on at $t=0$ for three different temperatures (in units of $\epsilon_0/k_B$). Bottom: final configuration for $T=0.0005$ exhibiting the quasi-conformal ``gravity rainbow" configuration.}
\label{fig.ConfExp}
\end{figure}

For the case of bulk superconducting samples, a possible way to induce a quasi-conformal vortex crystal is by compressing the system against a barrier using the transverse Lorentz force induced by a uniform current density applied parallel to the barrier. This concept is similar to the gravity rainbow experiment of Ref.~\cite{Pieranski1989}. In a superconductor, vortices trapped in strong pinning planes could act as barriers for the vortices in the weak pinning regions, as suggested in the cartoon of Fig.~\ref{fig.ConfExp}. For instance, vortex rows strongly pinned at twining planes, which are natural planar defects found in some high-$T_c$ crystals, are known to act as barriers for other vortices~\cite{Aprile1997}. Alternatively, one could fabricate strong pinning planes artificially by means of lithographic and irradiation techniques~\cite{Banerjee2003}. To test this idea, we performed Langevin dynamics simulations of vortices in a clean superconductor with a periodic array of identical pining potential planes modeled as $U_p(y)=-0.8\epsilon_0\exp(-y^2/2\xi^2)$, with $\xi=0.0625\lambda$. Two planes were placed at $y=0$ and $y=60\lambda$ and periodic boundary conditions were considered in a $120\lambda\times120\lambda$ simulation box containing 6000 vortices. After a simulated annealing procedure an essentially homogeneous triangular vortex lattice was obtained, but with slightly higher vortex density at the line positions. Then, we suddenly applied a current density $j=0.5\epsilon_0/\lambda\phi_0$ much smaller then the depinning current ($j_{dp}\simeq0.5\epsilon_0/\xi\phi_0$), thus guaranteeing a static final configuration of the vortices. The vortices were observed to quickly compress into a linear profile, while generating a large amount of defects, and thereupon healed progressively until finally stabilizing into a quasi-conformal configuration, as the one shown in Fig.~\ref{fig.ConfExp} (see also the Supplementary Video online). This self-organization process was observed to accelerate with the addition of small thermal fluctuations, which provide the defects with enhanced mobility allowing them to redistribute and eventually annihilate. The time evolution of the defect concentration for three different temperatures as well as the final configuration for $T=0.0005\epsilon_0/k_B$ are shown in Fig.~\ref{fig.ConfExp}, top panel. It is important to mention that in a real sample the healing of defects could be compromised by the ubiquitous material disorder, which tends to force the vortex system into a glassy state. However, in weak pinning superconductors, these states are characterized by a dilute distribution of topological defects that can be healed by e.g. thermal fluctuations, gentle ac shaking or a combination of both~\cite{Raes2014}. It is also noteworthy that, in an actual experiment, the vortex configuration observed at the sample surface might differ from that found in the bulk due to vortex bending, which is neglected in our 2D approach.

Finally, we stress that Eq.~\ref{eq.U} can be applied to other symmetries of the external confinement. For instance, in contrast to the case of unidirectional confinement considered in the present work, where only one conformal crystal is possible, a whole set of different conformal crystals could be stabilized by suitable choices of radial confinement. In this case, an accurate computation of the appropriate external potential is essential to discern the possible conformal structures. This opens an interesting perspective for discovering new crystalline phases of large nonuniform clusters of interacting particles.

\section*{Methods}

\subsection*{Langevin dynamics}

The vortex dynamics in superconductors is governed by overdamped motion, where the total force, $\bm{F}$, acting on the vortex is equilibrated by a viscous force, $\eta\bm{v}=\bm{F}$, with $\eta$ a viscous coefficient and $\bm{v}$ the vortex velocity. For a system of vortices interacting via a pair potential $V_\text{int}(\bm{r},\bm{r}')$ and subjected to an external potential $U(\bm{r})$, and also considering thermal fluctuations, the motion of the $i^{th}$ vortex is governed by the Langevin equation in the form
\begin{equation}    
  \eta\bm{v}_{i}=-\sum_{j}\bm{\nabla}V_{int}(\bm{r}_{i},\bm{r}_{j})-\bm{\nabla}U(\bm{r}_{i})+\bm{\Gamma}_{i}(T,t),
\label{M1}  
\end{equation}
where the sum in $j$ is over all the other vortices in the system and $\bm{\Gamma}_{i}(T,t)$ is the stochastic Langevin force representing the thermal kicks on the $i^{th}$ vortex, with $\langle \bm{\Gamma}_{i}(t) \rangle=0$ and $\langle \Gamma_{i,\alpha}(t)\Gamma_{j,\beta}(t') \rangle=2\eta k_{B}T\delta_{\alpha,\beta}\delta_{i,j}\delta(t-t')$. Here $\langle ... \rangle$ is the ensemble average, $k_{B}$ is the Boltzmann constant, and the  labels $\alpha, \beta$ indicate the components of the vector $\bm{\Gamma}$. 
The integration of the Eq.~(\ref{M1}) is numerically solved by the so-called stochastic Euler method, which approximates the time evolution of the vortex position by
\label{page35}
\begin{equation}
x_{\alpha}^{i}(t_{n+1})=x_{\alpha}^{i}(t_{n})+\frac{1}{\eta}f_{\alpha}^{i}(t_{n})dt+u(t_{n})\sqrt{(2k_{B}T/\eta)dt},
\label{Eq.4.49}
\end{equation} 
where $t_{n+1}=t_{n}+dt$, $f_{\alpha}^i=-\sum_{j}\frac{\partial}{\partial x_{\alpha}}V_{int}(\bm{r}_{i},\bm{r}_{j})-\frac{\partial}{\partial x_{\alpha}}U(\bm{r}_{i})$, and $u(t_{n})$ is a zero-mean, unit-variance Gaussian variable. In our simulations we used $dt=0.001 \eta\lambda^2/\epsilon_{0}$. In case of applied currents (Fig.~\ref{fig.ConfExp}), we add in Eq.~(\ref{M1}) the corresponding Lorentz force $\bm{F}_{L}=\phi_{0}\bm{j}\times\hat{z}$, where $\bm{j}$ is the density of current integrated along the sample thickness. In our model, the total current flow in the sample is given by the superposition of the applied current density $\bm{j}$ and the current induced by vortex density gradients $\bm{j}_v$, which induces a Lorentz force on a vortex $i$ given by the first term of Eq.~\eqref{Eq.4.49}. 

\subsection*{Simulated annealing}

In order to find the ground-state configuration, we performed the well-known simulated annealing procedure, where the vortex system is initialized in a random distribution at a temperature $T$ above the melting temperature, $T_{M}$, and is gently cooled down to $T=0$, where a low energy state is reached. After a time interval $\tau=200000dt$, we reduced the temperature following the rate $T=2T_{M}/m$, $m=1,2,3,\dots$. This cooling procedure was chosen as to guarantee that the system is equilibrated before each temperature step, while ensuring a high density of data points near the transition to the CVC phase. For $T<0.04 T_{M}$, we accelerated the cooling rate, once at this temperature range the system was found to be in the CVC state for all simulations.



\section*{Acknowledgements}

We thank L.R.E. Cabral for stimulating discussions and useful suggestions. This work was supported by the Brazilian Agencies FACEPE, under the grant Nos. APQ-2017-1.05/12 and APQ-0198-1.05/14, and CNPq.

\section*{Author contributions statement}

C.C.d.S.S. conceived and supervised the project and performed the analytical modeling. R.M.M. performed the numerical simulations and data analysis. The manuscript was written by C.C.d.S.S. with input from R.M.M.

\section*{Additional information}

\textbf{Competing financial interests:} The authors declare that they have no competing interests.

\end{document}